# Altering nodes types in controlling complex networks

Xizhe Zhang[1,2*], Yuyan Zhu[2,3] and Yongkang Zhao[2,4]

[1] *School of Computer Science and Engineering, Northeastern University, Shenyang, Liaoning, China*
[2] *Joint Laboratory of Artificial Intelligence and Precision Medicine of China Medical University and Northeastern University, Northeastern University, Shenyang, Liaoning, China*
[3] *The First Hospital of China Medical University, Shenyang, Liaoning, China*
[4] *National Institute of Health and Medical Big Data, China Medical University, Shenyang, Liaoning, China*
* E-mail: zhangxizhe@mail.neu.edu.cn

**Abstract** – Controlling a complex network towards a desired state is of great importance in many applications. A network can be controlled by inputting suitable external signals into some selected nodes, which are called driver nodes. Previous works found there exist two control modes in dense networks: distributed and centralized modes. For networks with the distributed mode, most of the nodes can be act as driver nodes; and those with the centralized mode, most of the nodes never be the driver nodes. Here we present an efficient algorithm to change the control type of nodes, from input nodes to redundant nodes, which is done by reversing edges of the network. We conclude four possible cases when reversing an edge and show the control mode can be changed by reversing very few in-edges of driver nodes. We evaluate the performance of our algorithm on both synthetic and real networks. The experimental results show that the control mode of a network can be easily changed by reversing a few elaborately selected edges, and the number of possible driver nodes is dramatically decreased. Our methods provide the ability to design the desired control modes of the network for different control scenarios, which may be used in many application regions.

## Introduction.

Controlling complex networks is a fundamental challenge in many complex systems [1-3]. A complex network can be driven to a desired state if some suitable external control signals are inputted into the network. Recently, the network control theory [3-5], which combine the power of control theory and network theory, are extensively used in analyzing complex networks of many application regions, e.g., identifying disease genes based on protein-protein interaction network [6-9], analyzing brain networks [10,11] and finding drug-target in metabolic network [12,13].

According to structural control theory, a network is said to be controllable if it can be driven from any initial state to a desired final state by inputting external control signals [2]. The driver nodes are the nodes which used to input external signals. Owing to the structural complexity of the network, we usually only need a few driver nodes to fully control the network. The minimum set of driver nodes (MDS) provides a useful perspective to understanding the control principles of complex networks [14-16].

Previous work [3] have found that the MDS of a network can be obtained by any maximum matching of the network, which the unmatched nodes are the driver nodes. Based on this framework, many works have been done to analyze the control properties of various networks. Liu et.al [3] found that the driver nodes tend to avoid hub noes and the degree distribution is closely related to the size of MDS of the networks. Menichetti et.al [17] further investigated the size of the MDS and found the number of the low in-degree nodes main determined the size of the MDS. Ruths et.al [18,19] classified the MDS into source nodes, external dilation points, internal dilation points and presented control profile to quantifies the control structures of complex networks. For control of the particular parts of the network, Piao et.al [20] presented a method which used immune nodes to facilitate the control of the communities. Gao et.al [21] presented an analytical framework to investigate the target control of complex networks.

For most of the real networks, the maximum matching is usually not unique, so does the MDSs. Because the number of MDSs may be numerous and exponential to network size [14], we can define two control types of nodes based on their participation in MDSs: 1. input nodes, which appear in at least one MDSs; 2. redundant nodes, which never appear in any MDSs. Previous works [16,22] found a surprising bifurcation phenomenon in dense networks, in which the majority of nodes are either input nodes or redundant nodes. This bimodality leads to two control modes: 1. centralized mode, which the most nodes of the network are redundant nodes; 2. distributed mode, which the most nodes of the network are input nodes. Our recent work [16] present the input graph, a simple geometry which revealing the complex control correlation of all nodes. We found that the giant components emerge in many real networks, which provides a clear topological explanation of bifurcation phenomenon emerging in dense networks. For the network with centralized control, only a few nodes can be inputted external control signals, that means the selection of control schemes are limited. For a network with distributed control, most nodes can be used as driver nodes and the control scheme is more flexible. Previous works [22] found that the control modes can be altered via structural perturbations, however, they did not provide the method to identify the minimum number of edges whose reversal can change the control mode of a network.

Although our previous works [16, 23] already present the method to alter control mode based on adding or removing edges, we believe it is still necessary to present a new method based on edge reversal. In some real control scenarios, reversing an edge may be more feasible than adding or removing edges. For example, in transportation network such as airline network, removing an edge may decrease the ability of the network, and adding an edge is too expensive. Therefore, reversing an edge may be a good choice to change the control properties of the network. Furthermore, to design the desired control scheme of a network, we may need all possible methods including adding, removing or reversing edges. Therefore, the method based on edge reversal is an important piece for design the control scheme of the complex network.

Here we present an efficient method to alter the control modes, which is mainly based on the edge reversal. The method is based on our previous works about the control connectivity between nodes, which prove that the input node must be reachable from at least one driver node. Therefore, to alter the control mode of a network, a simple way is to change the connectivity of the nodes. Based on this idea, we design an efficient method to alter a network from distributed control to centralized control. The experimental results on both synthetic and real networks showed that our method can be efficient alter the control mode of a network.

## Structural controllability and maximum matching

Consider a linear time-invariant networked systems G, its dynamics can be described by the following equation:

$$\frac{dx(t)}{dt} = Ax(t) + Bu(t) \tag{1}$$

where the state vector $x(t)=(x_1(t), …, x_N(t))^T$ denotes the value of $N$ nodes in the network at time $t$, $A$ is the transpose of the adjacency matrix of the network, $B$ is the input matrix that defines how control signals are inputted to the network, and $u(t)=(u_1(t), …, u_H(t))^T$ represents the $H$ input signals at time $t$.

According to the Kalman rank condition [2], the networked system $G$ is controllable if and only if the controllability matrix $C=(B, AB, A^2B, …, A^{N-1}B) \in \mathbb{R}^{N \times NM}$ has full rank, i.e., $rank(C)=N$. However, in some cases, the exact value of the nonzero elements in $A$ and $B$ is not available and the precise computation of $rank(C)$ is therefore unattainable. For those cases, Lin [1] introduced the weaker form of controllability, which is called structural controllability. The structural controllability theory treats $A$ and $B$ as structured matrices, i.e., their elements are either fixed zeros or free parameters. The system is structurally controllable if the maximum rank of $C$, denoted by $rank(C)$, can reach $N$ as a function of the free parameters in $A$ and $B$. Based on this framework, Liu et.al [3] present the minimum input theorem to identify the minimum number of driver nodes $N_D$ needed to control the whole network of $N$ nodes. They found that the unmatched nodes w.r.t. any maximum matching of the network is the *MDS*.

Consider the network representation $G(V, E)$, where $V$ is the set of nodes and E is the set of directed edges. To analyze the controllability of the directed network, we need to convert it to an undirected bipartite graph. The bipartite graph is built by splitting the node set $V$ into two node sets $V^{in}$ and $V^{out}$, where a node $n$ in $G$ is converted to two nodes $n^{in}$ and $n^{out}$ in $B$, and nodes $n^{in}$ and $n^{out}$ are, respectively, connected to the in-edges and out-edges of node $n$. Fig. 1 give an example of the network and its bipartite graph.

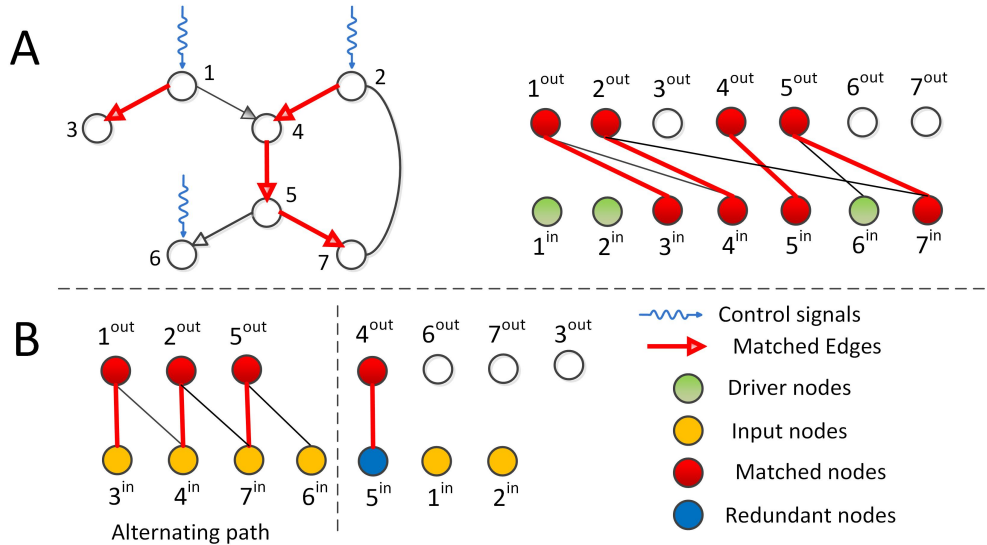

Fig. 1: Maximum matching and controllability of a network. a. a simple directed network and its corresponding bipartite graph representation. For any node $n$ in a directed network, it is converted to two nodes $n^{in}$ and $n^{out}$ in $B$, and nodes $n^{in}$ and $n^{out}$ are, respectively, connected to the in-edges and out-edges of node $n$. The red edges are maximum matching. The unmatched nodes of in-set are driver node. (b) Alternating paths and node classification. For any alternating paths start with driver nodes, all nodes within the paths are input

nodes, i.e., nodes {3,4,6,7}. The nodes which do not connect with driver nodes through alternating paths are redundant nodes, i.e., node {5}.

A matching is a set of edges in which no edge shares common nodes. The edges belong to a matching are called matched edges. A node is said to be matched if there is a matched edge linked to the node; Otherwise, the node is unmatched. The unmatched nodes in $V^{out}$ are called **unsaturated nodes**, and the unmatched nodes in $V^{in}$ are called **driver nodes**. A path is said to be an **alternating path** if the edges of the path are alternately in and not in the matching. An alternating path that begins and ends on the unmatched nodes is called the augmenting path. A maximum matching is a matching with the maximum number of edges. According to the structural controllability theory, for any maximum matching of the network, the set of unmatched nodes in $V^-$ is called Minimum Driver nodes Set (*MDS*).

The maximum matching of the network is not unique, so does the *MDS*. A node is called an **input node** if it appears in at least one *MDS*. Otherwise, we call it a **redundant node**. Previous works [16] found that in some dense networks, the majority of nodes of a network are either input nodes or redundant nodes. This bifurcation phenomenon leads to two control modes of the networks: centralized control and distributed control. For networks with distributed control, most of the nodes are input nodes. For the networks with centralized control, most of the nodes are redundant nodes. Therefore, to alter the control modes of a network, we need to change the control type of most of the nodes of a network.

## Method

In this section, we will introduce how to alter a network from distributed mode to centralized mode by reversing the direction of few selected edges. For a network with the distributed mode, most of the nodes are input nodes. Therefore, we only need to change these input nodes to redundant nodes. To change the control type of nodes, we first introduce two theorems of our previous works [16], which identify the input node and redundant node based on their connectivity with driver nodes.

**Theorem** 1: For any *MDS D* and a driver node $n \in D$, any node which can be reached by $n$ through any alternating path is an input node;

**Theorem** 2: For any *MDS D*, if node $m$ cannot be reached by any driver nodes of $D$ through any alternating path, $m$ must be a redundant node.

Based on the above two theorems, the control type of the nodes simply depends on their connectivity to the driver nodes. Thus, similar to the connected component of the graph theory, we can define the alternating connected components of the network, in which the nodes are connected with alternating paths. Based on their connectivity with driver nodes, we can define two types alternating connected components: 1. input component, which contains at least one driver node; and 2. matched component, which contains no driver node. Our previous works [16] proved that the control modes are rooted by the emerging of giant alternating connected components of complex networks. The networks with distributed mode have a giant input component, and the networks with centralized mode have a giant matched component. All nodes of the input component are input nodes and all those of the matched component are redundant nodes. Therefore, to alter the control mode of a network, we only need to change the type of the largest alternating connected component of the network.

Based on Theorem 1 and 2, our previous works [16,23] present the methods to change the control mode by adding edges or removing edges. In this paper, we consider how to change the network from distributed control to centralized control by reversing edges. Consider a network *G*, the basic idea of altering a network with a giant input component is to remove all driver nodes from the component. However, reversing edges are more complex than adding or removing edges. Suppose node $b$ is a driver node connected with input component *N* by edge $e(a,b)$, to reverse the direction of $e(a,b)$, we need to remove $e(a_{out}, b_{in})$ and add $e(b_{out}, a_{in})$ into the bipartite graph. Therefore, reversing an edge in the network equal to remove one edge and add another edge in the bipartite graph, which makes this problem more difficult than adding or removing edges. For a driver node $b$ and its in-edge $e(a, b)$, there are four possible cases in the bipartite graph after reversing the direction of $e(a, b)$:

**Case 1(Fig. 2B)**: node $a_{in}$ is not in the component *N* and $b_{out}$ is not connected with *N*;

**Case 2(Fig. 2C)**: node $a_{in}$ is in component *N* and $b_{out}$ is connected with *N*; both of them are matched nodes and cannot be reachable by any unmatched node through alternating path;

**Case 3(Fig. 2D)**: node $a_{in}$ is in component *N* and $b_{out}$ is connected with *N*; node $b_{out}$ can be reachable by at least one unmatched node through the alternating path, while node $a_{in}$ cannot.

**Case 4(Fig. 2E)**: node $a_{in}$ is in component *N* and $b_{out}$ is connected with *N*; node $a_{in}$ can be reachable by at least one unmatched node through the alternating path, while node $b_{out}$ cannot;

Note that node $a$ and $b$ cannot both in the component and be reachable by driver node, otherwise, there will be an augmenting path connected a and b, which is contradicting to the maximum matching. Fig. 2 show an example of these four cases. For cases 1, 2 and 3, it is easy to see, after reversing edge $e(a, b)$, the driver nodes $b$ is successfully detached from input component *N*. For case 4, however, node $a_{in}$ is still in the input component *N* and can be reached by driver node. Therefore, based on Theorem 1, all nodes of component *N* are still input nodes. In this case, we need to remove edge $e(b, a)$ to detach node $a$ from the input component *N*.

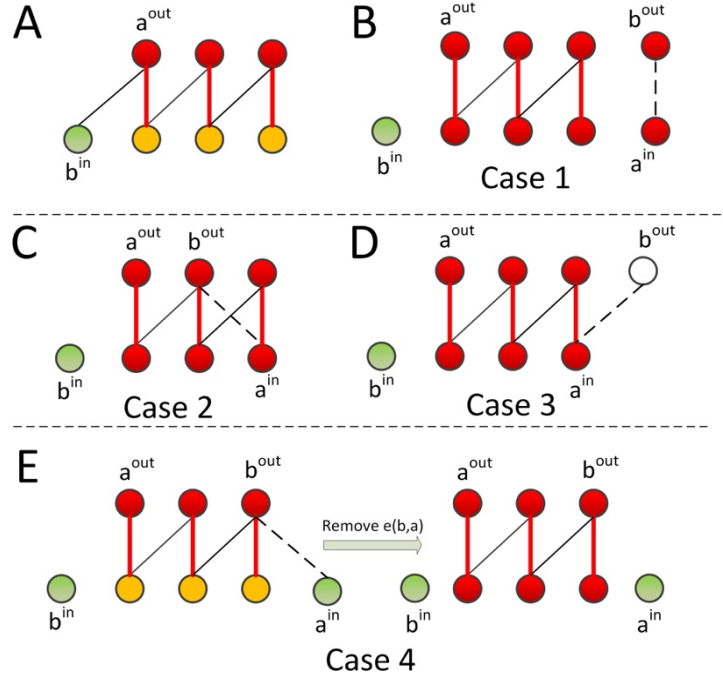

Fig. 2: Four possible cases when reversing an in-edge $e(a, b)$ of a driver node $b$. **A**. The input component $N$ before reversing the edge. Because node $b$ is a driver node, all nodes of $N$ are the input nodes; After reversing $e(a, b)$, we have four possible cases: **B.** (case 1) node $a_{in}$ and $b_{out}$ are detached from the component; **C.** (case 2): node $a$ and $b$ are both matched nodes and within component; **D.** (case 3): node $a_{in}$ is matched and $b_{out}$ is unmatched; **E.** (case 4): node $a_{in}$ is unmatched and $b_{out}$ is matched, in this case, if we add edge$(b, a)$, the component is still an input component, therefore, we need the removed $e(b, a)$.

In summary, to alter a network from distributed mode to centralized mode, we need to detach all driver nodes from the largest input component of the network. Therefore, the basic idea is to detach driver nodes by reversing the in-edges. We conclude four cases when reversing the in-edges of a driver node, in which the node is successfully detached from input component in cases 1,2 and 3. For case 4, we need to remove the in-edge of the input node. Based on the above discussion, the algorithm for altering control mode of the network are the follows:

Step1: Compute a maximum matching $M$ of the network $G$, and let the driver nodes be set $D$;

Step2: For each driver node $d \in D$, find all nodes of $V_{in}$ which connected to the driver node $d$ through alternating paths, denoted as $C(d)$;

Step3: For each pair of driver nodes $d_1$ and $d_2$, if $C(d_1) \cap C(d_2) \neq \varnothing$, merge the set $C(d_1)$ and $C(d_2)$, repeat it until no set can be merged;

Step4: Find the largest nodes set of Step 3, denote as set $N$, let the driver nodes of $N$ be the set $D_N$.

Step5: For each driver node $n \in D_N$, do the following:
   Step5.1: for each incoming edge $e(m,n)$ of node $n$, remove edge $e(m,n)$;
   Step5.2: if not ($m_{in} \in N$ and $m_{in} \in D$ and $n_{out}$ is connected to $N$), add edge $e(n,m)$.

In the above algorithm, we first obtain an *MDS* by computing any maximum matching of the network in Step 1 and then find the largest input component of the network in Steps 2 and 3. In Step 5, we first remove all in-edges of the driver nodes within the component and then add the inverted edges into the network except case 4. After processing all in-edges of driver nodes of the component, the nodes of the component will be turned into matched nodes and the network will be changed to the centralized mode.

## Experimental results

In this section, we evaluated the performance of our algorithm on both synthetic and real networks. First, we generated some scale-free networks with the number of nodes $N=10^4$, and the power exponent of degree distribution $\gamma_{in}=\gamma_{out}=3$. The average degree $k$ varies from 5 to 20 with increment 0.1. For each average degree $k$, we generated 20 random instances, in which the largest alternating components are input components. Therefore, there are total 3,200 network instances used to evaluate the methods.

To evaluate the performance of our algorithm, we first computed the percentage of input nodes *IN*. Fig. 3A showed the results of *IN*

versus average degree $k$ before reversing the edges. With the increase of the average degree $k$, the percentage of input nodes $IN$ increase from 0.25 to 1, indicating that denser networks have clearly distributed control mode. Furthermore, with the increase of the average degree, the distributed mode of the networks is emerging, which is consisted with previous works [22]. For the sparse networks with low $k \in [5,9]$, $IN$ are range from 0.25 to 0.4, which means that these networks have no clearly control mode [22]. For the dense networks with high average degree $k \in [9,20]$, the input nodes are the majority of the networks, which means there exist a giant input component in the networks.

Next, we computed the percentage of input nodes after reversing the edges. Fig. 3B showed the percentage of input nodes before and after edge reversal, denoted as $IN_{before}$ and $IN_{after}$, respectively. The results showed that $IN_{after}$ is decreased after reversing edges for all network instances. Furthermore, for dense networks which $k \in [10,20]$, $IN_{after}$ are significantly decreased, i.e., from nearly 90% to 5% when $k=14$. Fig. 4A showed the percentage of altered input nodes $IN_{after}/IN_{before}$ for different $k$. we can see that for the dense network which $k>10$, more than 80% input nodes are changed after reversing edges. However, for the networks with low average degree, the input nodes are not significantly changed, as Fig. 3B showed. For networks which $k<8$, the percentage of input nodes only changed about 10% after reserving edges. This phenomenon is rooted that these networks do not have a giant input component, therefore, they are neither distributed mode or centralized mode.

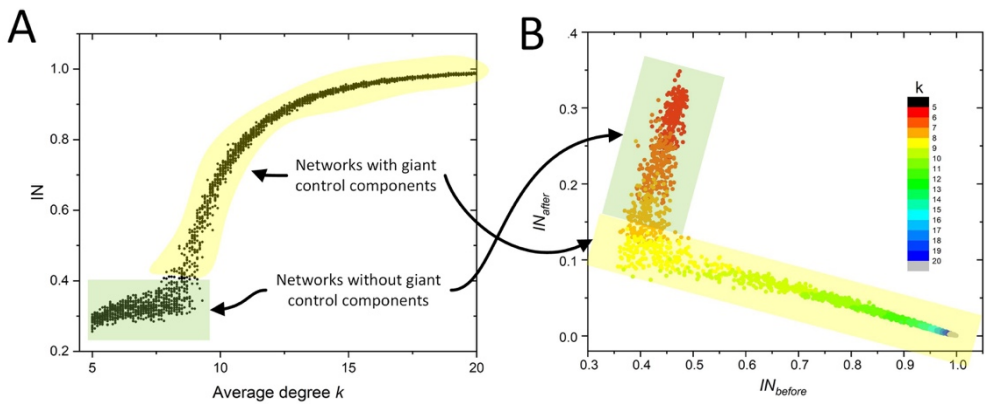

Fig. 3: Performance of our algorithm. We generated 3,500 network instances with $k \in [5,20]$. Each dot represents a network instance. **A**. The percentage of input nodes $IN$ versus average degree $k$. **B**. The percentage of input nodes before and after reversing edges.

Next, we computed the number of reversed and removed edges for altering the control mode of the network. Fig. 4 showed that for most networks, the input nodes are significantly decreased while very few edges need to be modified. For the network with $k>10$, only 1% edges are modified, and nearly 90% input nodes are changed, which show our algorithm are very efficient to change the control mode of the dense networks. We also compute the fraction of reversed edges and removed edges when altering the control mode of the network. Fig. 4B showed that with the increase of average degree, we need few edges to change the control modes of the network. Furthermore, for all network instances, more than 94% of modified edges are reversed edges, while only very few edges are needed to be removed. For dense networks with $k>10$, most networks do not need removed edges to alter the control mode. Therefore, our algorithm only needs to remove very few edges when altering control mode.

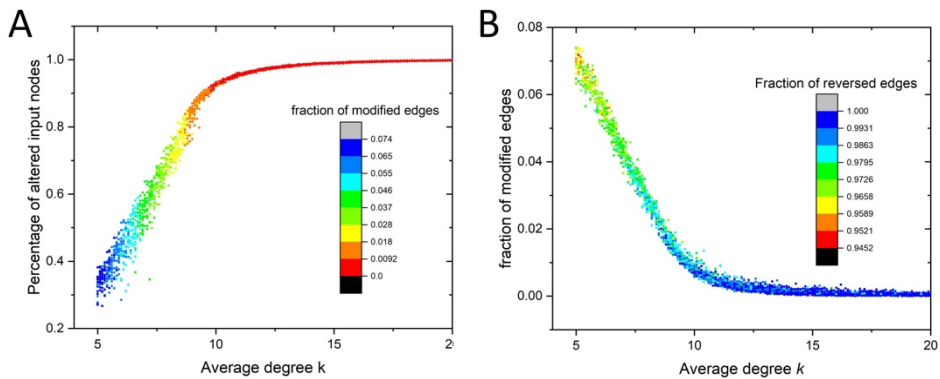

Fig. 4: The modified edges versus average degree $k$. **A.** the percentage of altered input nodes are significantly increased with the average degree $k$, while only need very few edges to be modified; **B.** the modified edges are decreased with the average degree $k$, in which most edges are the reversed edges.

To demonstrate the feasibility of our algorithm, we applied our algorithms to several real networks. These networks are selected based on their diversity of topological structure, including biological networks, social networks, and technical networks. For each network, we show its type, name, number of nodes ($N$) and edges ($L$), size of the largest input component $IC_{max}$, the percentage of modified edges ($p_m=Num_{me}/L$), the percentage of reversed edges ($p_r=Num_{re}/Num_{me}$), the percentage of changed input nodes ($\Delta n_D$) and the percentage of size of changed input component $\Delta IC_{max}$. The detailed results are listed in Table.1. Overall, the largest input components of these networks can be completed altered by reversing and remover edges. For the high average degree network, such as Amazon0505 product co-purchasing network, 87.83% of input nodes have dramatically changed after reversing less than 1% edges. Similar results are found in the Twitter network and Slashdot network. For the other networks such as P2P networks, we can still change the control modes of the network, yet the number of edges needs to be modified are relatively high. This may be caused by these networks did not have a dense connected input component. However, compared previous works which reversing all edges to change the mode, our algorithm still has better performance.

Table 1: Result of some real networks. For each network, we show its type, name, number of nodes ($N$) and edges ($L$), size of the largest input component $IC_{max}$, the percentage of modified edges ($p_m=Num_{me}/L$), the percentage of reversed edges ($p_r=Num_{re}/Num_{me}$), the percentage of changed input nodes ($\Delta n_D$) and the percentage of size of changed input component $\Delta IC_{max}$

| Type | Name | N | L | ICmax | pm | pr | ΔnD | ΔICmax |
|---|---|---|---|---|---|---|---|---|
| Food Web | Mangrove | 97 | 1492 | 55.67% | 5.03% | 58.67% | 37.11% | 98.15% |
| | Silwood | 154 | 370 | 84.42% | 75.14% | 80.94% | 25.97% | 96.92% |
| Trust | Slashdot0902 | 82168 | 948464 | 91.23% | 9.41% | 80.92% | 86.73% | 99.99% |
| Citation | ArXiv-HepTh | 27770 | 352807 | 14.44% | 0.60% | 85.29% | 10.90% | 98.98% |
| WWW | NotreDame | 325729 | 1497134 | 53.90% | 15.20% | 86.36% | 14.71% | 94.53% |
| | Google | 875713 | 5105039 | 21.81% | 4.21% | 92.12% | 11.41% | 99.72% |
| Internet | p2p-1 | 10876 | 39994 | 90.58% | 50.08% | 78.79% | 52.65% | 99.84% |
| | p2p-2 | 8846 | 31839 | 90.55% | 51.18% | 77.55% | 53.57% | 99.79% |
| | p2p-3 | 8717 | 31525 | 91.75% | 50.80% | 76.92% | 53.60% | 99.30% |
| Organizational | Consulting | 46 | 879 | 97.83% | 3.53% | 93.55% | 93.48% | 97.78% |
| Social communication | UClonline | 1899 | 20296 | 79.94% | 12.36% | 97.81% | 50.45% | 99.93% |
| Product co-purchasing networks | Amazon0505 | 410236 | 3356824 | 91.35% | 0.92% | 94.64% | 87.83% | 99.99% |
| | Amazon0302 | 262111 | 1234877 | 8.76% | 0.32% | 95.00% | 8.07% | 97.11% |
| Social network | twitter_combined | 81306 | 1768149 | 79.40% | 3.04% | 90.66% | 60.69% | 99.75% |

## Conclusion

Controlling a complex network is a fundamentals task for various applications. The complex structure of these networks forming two distinct control modes of the networks, which may be useful for different control purposes. The control type of nodes has been proved useful in identifying drug target [13], cancer genes [24,25] and understanding the role of neurons [26]. Understanding and manipulating control properties of the networks are of great importance in many applications, especially biological networks.

Here we present a method to alter the control modes of the networks by reversing very few edges. The results showed that we can efficiently change a network from a distributed mode to centralized mode. This may be useful in some control scenario. For example, consider a network with distributed control modes, which means there exist many available control schemes can control the network, which may provide weakness for some malicious attackers. A feasible response is to change the control mode to centralized mode, which means only very few driver nodes can control the network.

Our algorithm can also be used to change the control type of specific node of the network. For example, to change a node from input nodes to redundant nodes, we can simply detach the nodes from the input component or alter the mode of the input component. Previous works [9] have found that indispensable nodes of Protein-Protein interaction networks may be critical in cancer study, and indispensable nodes are parts of redundant nodes. Therefore, the type change of these driver genes may be useful in the transitions between health and disease state [9].

The limitation of our algorithm is that it can only change the control modes from distributed mode to centralized mode. To change the mode from centralized to distributed mode are more complex and we will finish it in the future work.

## Conflicts of interest

The authors declare no conflicts of interests.